\documentclass{aa}
\usepackage{epsf}

\begin{document}

\thesaurus{05 (08.11.1; 08.06.2; 08.16.5; 08.12.1; 05.01.1)}

\title{Kinematics of T~Tauri stars in Chamaeleon}

\author{Sabine Frink \inst{1} \and Siegfried R\"oser \inst{1} \and
 Juan M.\ Alcal\'a \inst{2,3} \and Elvira Covino \inst{2} \and Wolfgang 
 Brandner \inst{4}}
\offprints{Sabine Frink, e-mail: sabine@ari.uni-heidelberg.de}
\institute{Astronomisches Rechen-Institut Heidelberg,
    M\"onchhofstra{\ss}e~12-14, D-69120 Heidelberg, Germany \and
    Osservatorio Astronomico di Capodimonte, Via Moiariello~16,
    I-80131 Napoli, Italy \and
    Instituto Nacional de Astrof\'{\i}sica, Optica y Electr\'onica, A.P. 51 y 216
    C.P. 72000, Puebla, M\'exico \and
    Caltech - JPL/IPAC, Mail Code 100-22, Pasadena, CA 91125, USA}
\date{Received / Accepted}

\maketitle

\begin{abstract}
We study the kinematics of T~Tauri stars (TTS) located in the cores of the
Chamaeleon clouds as well as far off these clouds. Our sample comprises
2~early type stars known to be related to Cha~{\sc i}, 6 classical
(CTTS) and 6 weak-line T~Tauri stars (WTTS) known before the ROSAT
mission, and 8 bona-fide pre-main sequence (PMS) stars as well as 23
presumably older stars discovered with ROSAT (Alcal\'a et al.\ 1995;
Covino et al.\ 1997). Altogether we present
proper motions for 45~stars, taken from the Hipparcos, ACT and STARNET
catalogues. For 12~stars of our sample parallaxes measured by
Hipparcos are available, and we use them to derive constraints on the
distance distribution of the other stars in our sample.
Our analysis of the proper motions allows us to divide the sample 
into several subgroups.

We analyse the motions of the stars in connection with different star
formation scenarios and find them consistent with both the high
velocity cloud (HVC) impact model (L\'epine \& Duvert 1994) and
the cloudlet model (Feigelson 1996), whereas the data seem to be
inconsistent with any kind of a dynamical ejection model.

\end{abstract}

\section{Introduction}
The Chamaeleon cloud complex, located in the southern hemisphere, was first
discussed as a separate system of dark clouds by Hoffmeister (1962). He 
identified 26~RW Auriga type variable stars in this region, 
some of which also showed H$\alpha$-emission.

\begin{table*}
\begin{center}
\caption[]{\label{liste1} Stars which could be identified either in the
Hipparcos (HIP), ACT (A) or STARNET (S) catalogue and which were known to be 
associated with the Cha~{\sc i} or the Cha~{\sc ii} cloud before the ROSAT 
mission. Additional designations for the stars are given in the last column.
Please note that the mean errors of the proper motions in right ascension are
given with the factor $\cos \delta$.}
\begin{tabular}{@{}lr@{\ }r
                r@{}c@{}r@{}c@{}r@{}c@{}l
                r@{}c@{}r@{}c@{}r@{}c@{}l
                rrrrr@{\,}ll@{}}
\hline\noalign{\smallskip}
& \multicolumn{2}{c}{HIP No./} & \multicolumn{7}{c}{RA} &
\multicolumn{7}{c}{DEC} & 
\multicolumn{1}{c}{$\mu_{\alpha}$} 
& \multicolumn{1}{c}{$\mu_{\delta}$} & 
\multicolumn{1}{l}{\raisebox{1ex}[-1ex]{$\sigma_{\mu_{\alpha}}$}
\raisebox{-1ex}[1ex]{\scriptsize \hspace*{-0.6cm}$\cos\delta$}} &
\multicolumn{1}{c}{$\sigma_{\mu_{\delta}}$} & \multicolumn{2}{c}{$d$} & \\
\raisebox{1.5ex}[-1.5ex]{object} & 
\multicolumn{2}{c}{GSC No.} & 
\multicolumn{7}{c}{($\alpha_{J2000.0}$)} & 
\multicolumn{7}{c}{($\delta_{J2000.0})$} &
\multicolumn{2}{c}{[mas\,yr$^{-1}$]} & 
\multicolumn{2}{c}{[mas\,yr$^{-1}$]} & \multicolumn{2}{c}{[pc]} &
\raisebox{1.5ex}[-1.5ex]{other designations} \\
\noalign{\smallskip}\hline\hline\noalign{\smallskip}
\multicolumn{24}{l}{early type stars} \\
\noalign{\smallskip}\hline\noalign{\smallskip}
HD 97048 & \multicolumn{2}{c}{HIP 54413} & 11 &$^{h}$& 8 &$^{m}$&  3.&$\!\!^{s}$& 32 & -77 &$^{\mbox{\scriptsize o}}$& 39 &\arcmin& 17. &$\!\!\arcsec$& 5 & -92 &   2 & 0.8 & 0.8 & 175 & $^{+27}_{-20}$ & CHXR 29, Sz 25, PPM 370994 \\
HD 97300 & \multicolumn{2}{c}{HIP 54557} & 11 & & 9 & & 50.& & 02 & -76 & & 36 & & 47.& & 7 & -94 &  -1 & 1.0 & 0.9 & 188 & $^{+43}_{-30}$ & CHXR 42, PPM 371004, \\
\multicolumn{24}{r}{A 9410 2805} \\
\noalign{\smallskip}\hline\noalign{\smallskip}
\multicolumn{24}{l}{classical T Tauri stars} \\
\noalign{\smallskip}\hline\noalign{\smallskip}
Sz 6   &\multicolumn{2}{c}{HIP 53691}& 10 &$^{h}$ & 59 &$^{m}$ &  6. &$\!\!^{s}$ & 97 & -77 &$^{\mbox{\scriptsize o}}$ &  1 &\arcmin & 40. &$\!\!\arcsec$ & 3 & -97  &   2 & 2.1 & 1.8 & 143 & $^{+53}_{-30}$ & CHXR 6, AS 9414 186 \\
CS Cha &  S 9414 &  574 & 11 & &  2 & & 24. & & 79 & -77 & & 33 & & 35. & & 4 & -160 &  16 & 6.7 & 6.7 & & & CHXR 10, Sz 9 \\
Sz 19  &\multicolumn{2}{c}{HIP 54365}& 11 & &  7 & & 20. & & 72 & -77 & & 38 & &  7. & & 3 & -113 &  3 & 3.1 & 2.9 & 210 & $^{+303}_{-78}$ & CHXR 23, AS 9414 743 \\
VW Cha &  S 9414 &  754 & 11 & &  8 & &  1. & & 25 & -77 & & 42 & & 28. & & 6 &  -218 &   8 & 7.7 & 7.7 & & & CHXR 31, Sz 24 \\
CV Cha $^{(*)}$ \hspace*{-0.7cm} &  S 9410 &   60 & 11 & & 12 & & 27. & & 75 & -76 & & 44 & & 22. & & 4 &  -107 &   4 & 4.4 & 4.4 & & & CHXR 51, Sz 42, A 9410 60 \\
BF Cha &  S 9417 &  708 & 13 & &  5 & & 20. & & 57 & -77 & & 39 & &  1. & & 6 &   -98 &   1 & 4.5 & 4.5 & & & Sz 54 \\
\noalign{\smallskip}\hline\noalign{\smallskip}
\multicolumn{23}{l}{weak-line T Tauri stars} \\
\noalign{\smallskip}\hline\noalign{\smallskip}
CHXR 8  & A 9414 &  444 & 11 &$^{h}$ &  0 &$^{m}$ & 14. &$\!\!^{s}$ & 50 & -77 &$^{\mbox{\scriptsize o}}$ & 14 &\arcmin & 38. &$\!\!\arcsec$ & 0 & -114 &  13 & 2.6 & 3.7 & & & S 9414 444 \\
CHXR 11 & S 9414 &  642 & 11 & &  3 & & 11. & & 45 & -77 & & 21 & &  3. & & 7 & -166 &  20 & 7.4 & 7.4 & & & \\
CHXR 32 & S 9414 &  640 & 11 & &  8 & & 14. & & 79 & -77 & & 33 & & 52. & & 1 & -260 &  34 & 7.7 & 7.7 & & & Glass I \\
Sz 41 & S 9410 &  300 & 11 & & 12 & & 26. & & 10 & -76 & & 37 & &  3. & & 7 & 230 &  44 & 7.7 & 7.7 & & & CHXR 50 \\
CHXR 56 & S 9414 &  209 & 11 & & 12 & & 42. & & 50 & -77 & & 22 & & 25. & & 9 & -195 & -30 & 7.1 & 7.1 & & & HM Anon \\
T Cha &\multicolumn{2}{c}{HIP 58285}& 11 & & 57 & & 13. & & 53 & -79 & & 21 & & 31. & & 6 & -215 & -10 & 5.1 & 3.8 & 66 & $^{+19}_{-12}$ & RXJ 1157.2-7921, AS 9419 1187\\
\noalign{\smallskip}\hline\\[0.8ex]
\end{tabular}
\begin{tabular}{c@{\,\,}p{17cm}}
$^{(*)}$ & CV~Cha is a double system (CCDM 11125-7644) with 2 entries in 
Hipparcos (HIP 54738 and HIP 54744) and an orbital solution qualified as poor.
Its parallax is rather uncertain (3.14$\pm$7.39\,mas).
\end{tabular}
\end{center}
\end{table*}

\begin{table*}
\begin{center}
\caption[]{\label{liste2} Stars with proper motions from the
ROSAT sample investigated by Alcal\'a et al.\ (1995, 1997) and C97. 
The data are again taken from the Hipparcos, ACT and STARNET catalogues.
The classification in T Tauri stars and ZAMS and other stars is based
on the lithium criterium as applied by C97. Note however that some of the
stars classified as ZAMS stars or stars of unknown nature fall well above
the main sequence when placing them in the HR diagram with the help of
the Hipparcos parallax (Neuh\"auser \& Brandner 1998).}
\begin{tabular}{@{}lr@{\ }r
                r@{}c@{}r@{}c@{}r@{}c@{}l
                r@{}c@{}r@{}c@{}r@{}c@{}l
                rrrrr@{\,}ll@{}}
\hline\noalign{\smallskip}
object & \multicolumn{2}{c}{HIP No./} & \multicolumn{7}{c}{RA} &
\multicolumn{7}{c}{DEC} &
\multicolumn{1}{c}{$\mu_{\alpha}$} 
& \multicolumn{1}{c}{$\mu_{\delta}$} & 
\multicolumn{1}{l}{\raisebox{1ex}[-1ex]{$\sigma_{\mu_{\alpha}}$}
\raisebox{-1ex}[1ex]{\scriptsize \hspace*{-0.6cm}$\cos\delta$}} &
\multicolumn{1}{c}{$\sigma_{\mu_{\delta}}$} & \multicolumn{2}{c}{$d$} & \\
RXJ ... &
\multicolumn{2}{c}{GSC No.} & 
\multicolumn{7}{c}{($\alpha_{J2000.0}$)} & 
\multicolumn{7}{c}{($\delta_{J2000.0})$} &
\multicolumn{2}{c}{[mas\,yr$^{-1}$]} & 
\multicolumn{2}{c}{[mas\,yr$^{-1}$]} & \multicolumn{2}{c}{[pc]} &
\raisebox{1.5ex}[-1.5ex]{other designations} \\
\noalign{\smallskip}\hline\hline\noalign{\smallskip}
\multicolumn{23}{l}{T Tauri stars} \\
\noalign{\smallskip}\hline\noalign{\smallskip}
0837.0-7856  & A 9402 &  921 &  8 &$^{h}$ & 36 &$^{m}$ & 56. &$\!\!^{s}$ & 24 & -78 &$^{\mbox{\scriptsize o}}$ & 56 &\arcmin & 45.&$\!\!\arcsec$ & 7 & -147 &  26 & 1.4 & 1.8 & & & S 9402 921 \\
0850.1-7554  & A 9395 & 2139 &  8 & & 50 & &  5. & & 44 & -75 & & 54 & & 38.& & 2 &   -79 &  31 & 2.2 & 1.4 & & & S 9395 2139 \\
0951.9-7901  & A 9404 &  195 &  9 & & 51 & & 50. & & 68 & -79 & &  1 & & 37.& & 8 &  -149 &  39 & 0.9 & 2.6 & & & PPM 370508, S 9404 195 \\
1150.4-7704  & S 9415 & 1685 & 11 & & 50 & & 28. & & 23 & -77 & &  4 & & 38.& & 4 &  -223 & -16 & 4.5 & 4.5 & & & \\
1158.5-7754a &\multicolumn{2}{c}{HIP 58400} & 11 & & 58 & & 28. & & 15 & -77 & & 54 & & 29.& & 6 &  -198 &  -1 & 1.4 & 1.1 & 86 & $^{+11}_{-9}$ & AS 9415 1238 \\
1159.7-7601  &\multicolumn{2}{c}{HIP 58490} & 11 & & 59 & & 42. & & 27 & -76 & &  1 & & 26.& & 1 &  -165 &  -5 & 1.7 & 1.5 & 92 & $^{+17}_{-13}$ & AS 9411 2191 \\
1201.7-7859  & A 9420 & 1420 & 12 & &  1 & & 39. & & 13 & -78 & & 59 & & 16.& & 9 & -213 &  -5 & 0.8 & 0.8 & & & PPM 785565, S 9420 1420 \\
1239.4-7502  & A 9412 &   59 & 12 & & 39 & & 21. & & 27 & -75 & &  2 & & 39.& & 2 & -159 & -11 & 2.6 & 3.1 & & & S 9412 59\\
\noalign{\smallskip}\hline\noalign{\smallskip}
\multicolumn{23}{l}{ZAMS and other stars} \\
\noalign{\smallskip}\hline\noalign{\smallskip}
0849.2-7735  & A 9399 & 1491 &  8 &$^{h}$ & 49 &$^{m}$ & 11. &$\!\!^{s}$ & 10 & -77 &$^{\mbox{\scriptsize o}}$ & 35 &\arcmin & 58.&$\!\!\arcsec$ & 6 & -36 &  19 &  2.1 & 2.6 & & & PPM 370092, S 9399 1491 \\
0853.1-8244  & S 9506 & 1465 &  8 & & 53 & &  5. & & 26 & -82 & & 44 & &  0.& & 4 & -2 & -21 & 3.7 & 3.7 & & & \\
0917.2-7744  & A 9399 & 2104 &  9 & & 17 & & 10. & & 38 & -77 & & 44 & &  2.& & 0 & -159 & 13 & 2.1 & 0.8 & & & S 9399 2104\\
0919.4-7738N &\multicolumn{2}{c}{HIP 45734} &9 & & 19 & & 24. & & 67 & -77 & & 38 & & 36.& & 4 & -502 &  70 & 1.3 & 1.1 & 73 & $^{+7}_{-6}$ & PPM 370271 \\
0928.5-7815  & A 9400 & 1990 &  9 & & 28 & & 15. & & 02 & -78 & & 15 & & 22.& & 4 & -118 &  17 & 2.2 & 2.5 & & & PPM 370343 \\
0936.3-7820  &\multicolumn{2}{c}{HIP 47135} &  9 & & 36 & & 17. & & 82 & -78 & & 20 & & 41.& & 6 & -361 &  50 & 0.7 & 0.6 & 63 & $^{+3}_{-3}$ & PPM 370394, AS 9400 1713  \\
0952.7-7933  & A 9404 & 1702 &  9 & & 53 & & 13. & & 73 & -79 & & 33 & & 28.& & 4 & -82 &   6 & 1.6 & 2.1 & & & PPM 370518, S 9404 1702 \\
1009.6-8105  & A 9409 & 1040 & 10 & &  9 & & 58. & & 30 & -81 & &  4 & & 11.& & 4 &  86 & -24 & 2.4 & 0.8 & & & PPM 377161, S 9409 1040 \\
1039.5-7538S &\multicolumn{2}{c}{HIP 52172}& 10 & & 39 & & 31. & & 73 & -75 & & 37 & & 56. & & 3 & 13 & 17 & 1.6 & 1.4 & 128 & $^{+29}_{-20}$ & \\
1120.3-7828  & S 9415 & 2314 & 11 & & 20 & & 19. & & 68 & -78 & & 28 & & 21.& & 0 &  46 &  73 & 3.7 & 3.7 & & & \\
1125.8-8456  &\multicolumn{2}{c}{HIP 55746} & 11 & & 25 & & 17. & & 74 & -84 & & 57 & & 16.& & 3 & -545 &  12 & 0.7 & 0.6 & 83 & $^{+4}_{-4}$ & PPM 377341, AS 9511 1593 \\
1140.3-8321  & S 9507 & 2466 & 11 & & 40 & & 16. & & 52 & -83 & & 21 & &  0.& & 3 & -366 &  28 & 3.9 & 3.9 & & & \\
1203.7-8129  & S 9424 &  988 & 12 & &  3 & & 24. & & 66 & -81 & & 29 & & 55.& & 3 &   -98 & -10 & 3.7 & 3.7 & & & \\
1207.9-7555  & A 9412 & 2105 & 12 & &  7 & & 51. & & 16 & -75 & & 55 & & 16.& & 1 &  -639 &  -7 & 3.1 & 0.8 & & & PPM 785598, S 9412 2105 \\
1209.8-7344  & S 9239 & 1321 & 12 & &  9 & & 42. & & 79 & -73 & & 44 & & 41.& & 5 &   -51 &  -5 & 3.6 & 3.6 & & & \\
1217.4-8035  & A 9420 &  439 & 12 & & 17 & & 26. & & 90 & -80 & & 35 & &  6.& & 9 &    -5 & -11 & 2.7 & 2.3 & & & PPM 377437, S 9420 439 \\
1220.6-7539  & A 9412 & 1370 & 12 & & 20 & & 34. & & 37 & -75 & & 39 & & 28.& & 7 &  -472 &   3 & 1.5 & 2.8 & & & PPM 785640, S 9412 1370 \\
1223.5-7740  & A 9416 &  555 & 12 & & 23 & & 29. & & 04 & -77 & & 40 & & 51.& & 4 &  -306 &  12 & 0.8 & 0.8 & & & PPM 371482, S 9416 555 \\
1225.3-7857  & A 9420 &  742 & 12 & & 25 & & 13. & & 42 & -78 & & 57 & & 34.& & 8 &  -123 & -23 & 4.2 & 1.0 & & & PPM 371498, S 9420 742 \\
1233.5-7523  &\multicolumn{2}{c}{HIP 61284} & 12 & & 33 & & 29. & & 78 & -75 & & 23 & & 11.& & 3 & -370 &  13 & 1.1 & 1.1 & 66 & $^{+5}_{-4}$ & PPM 371552, AS 9412 190 \\
1307.3-7602  & A 9413 & 2147 & 13 & &  7 & & 22. & & 92 & -76 & &  2 & & 36.& & 2 &   -57 &   7 & 0.8 & 2.0 & & & PPM 785854, S 9413 2147 \\
1325.7-7955  & S 9434 &   97 & 13 & & 25 & & 41. & & 79 & -79 & & 55 & & 16.& & 2 &   -50 &   0 & 5.0 & 5.0 & & & \\
1349.2-7549E & A 9426 &  682 & 13 & & 49 & & 12. & & 92 & -75 & & 49 & & 47.& & 5 &  -267 & -31 & 1.9 & 2.1 & & & PPM 372040, S 9426 682 \\
\noalign{\smallskip}\hline\\[0.8ex]
\end{tabular}
\end{center}
\end{table*}

Objective prism surveys conducted in the following decades increased the number
of emission-line stars suspected to be associated with the Chamaeleon dark
clouds. First results were reported by Henize (1963).
The surveys conducted in 1962 and 1970 revealed 32 emission-line
stars (Henize \& Mendoza 1973),
which were all confirmed in the extensive survey by Schwartz (1977) in
the southern hemisphere and, particularly, in the Chamaeleon region. Altogether
he found 45~stars in the Cha~{\sc i} cloud and 19 in the Cha~{\sc ii} cloud.
In another objective prism survey Hartigan (1993) found 
21~new H$\alpha$ emission-line objects in Cha~{\sc i} and 5 in Cha~{\sc ii}.

Gregorio-Hetem et al.\ (1992) and Torres et al.\ (1995) used far-infrared
IRAS colours to preselect T~Tauri star (TTS) candidates over the whole sky 
and found, among others, 8 bona-fide plus 1 probable TTS in or around the 
Chamaeleon region.

Parallel to the objective prism surveys X-ray surveys have expanded the
membership lists since the late eighties. X-ray observations with the Einstein
Observatory revealed 22~X-ray sources, of which 6 or 7  were
associated with new probable cloud members (Feigelson \& Kriss 1989).
By means of high dispersion optical spectroscopy, Walter (1992) confirmed
the pre-main sequence (PMS) nature for 5 of
these sources as well as for 2~new candidates.

The ROSAT All-Sky Survey (RASS) has revealed 179 X-ray sources
in total, of which 77 have been classified as WTTS (Alcal\'a et al.\ 1995). 
They are
located not only near the known cloud structures, but also up to 10~degrees 
away from any known cloud material. For about 70 of them high resolution
spectroscopy is now available, and more than 50\% of the sources turn
out to be in fact very young weak-line T~Tauri stars (Covino et al.\ 1997, C97).
Some additional sources were found from ROSAT pointed observations in the
Cha~{\sc i} cloud (Feigelson et al.\ 1993).

Altogether, the membership list compiled by Lawson et al.\ (1996) contains 
117~bona-fide or probable T~Tauri stars in the inner region of the association,
apart from the wider distributed population investigated by C97. 

The discovery of many weak-line T~Tauri stars up to about 50 pc
away from the known molecular cloud cores of several nearby star forming
regions (SFR) (e.g.\ 
Cha\-mae\-le\-on: Alcal\'a et al.\ 1995; 
Orion: Alcal\'a et al.\ 1996;
Lupus: Krautter et al.\ 1997, Wichmann et al.\ 1997b; 
Taurus-Auriga: Neuh\"auser et al.\ 1997, Magazz\'u et al.\ 1997) 
has raised the question about their origin.
Before, with the exception of TW Hya (Ruci\'nski \& Krautter 1983),
pre-main sequence stars had only been found near the densest parts of molecular
clouds, and it was assumed that all stars originate from these cloud
cores. 
While Wichmann et al.\ (1997b) found that the mean age of WTTS far from the
clouds was higher than for WTTS projected onto the dark clouds in Lupus, 
Alcal\'a et al.\ (1997) found some of the youngest WTTS far from the
molecular clouds in Chamaeleon.
In order to travel 30\,pc in $5\cdot10^{6}$\,yrs 
(a typical T~Tauri age in the Cha\,{\sc i} cloud)   
a relative velocity of about
6\,km\,s$^{-1}$ would be required, much more than the value of 
1-2\,km\,s$^{-1}$ considered typical for the intrinsic velocity dispersion
by Jones \& Herbig (1979) or the value of $0.9\pm0.3$\,km\,s$^{-1}$ derived
by Dubath et al.\ (1996) using the radial velocities of 10~stars associated
with the Cha\,{\sc i} cloud.

Several scenarios have been put forward to account for the widely spread
population of WTTS, including models where star formation takes place in the
cloud cores and the stars are ejected out of these clouds subsequently
(Sterzik \& Durisen 1995) as well as models where star formation takes place in
small cloudlets which disappear after the formation process (Feigelson 1996).

The kinematic signature of these processes should still be visible: while in
the first scenario the velocity vectors of the stars should point away from the
dense cores from where they were ejected, the second scenario may have produced
small numbers of comoving WTTS with rather high relative velocities between
different groups.

Triggered star formation by means of supernova explosions or the
impacts of high velocity clouds (HVC) with the galactic plane have been
proposed to explain the positions of some SFR with respect to the galactic plane
(Tenorio-Tagle et al.\ 1987, L\'epine \& Duvert 1994). 
Nevertheless, in Chamaeleon there is no evidence of any OB association which
could have triggered star formation. L\'epine \& Duvert however successfully
modeled the observed geometry of the clouds with respect to the galactic
plane with a rather simple model of a high velocity cloud impact, which also
may have given rise to the observed widely spread PMS stars.

In this paper we analyse the kinematics of these stars in terms of the above
models. Proper motions are taken from the Hipparcos (ESA 1997), and ACT 
(Urban et al.\ 1997) catalogues as well as from STARNET (R\"oser 1996), which
gives proper motions for about 4.3~million stars with an accuracy of about 
5~mas\,yr$^{-1}$ and is a database well suited to study this population of 
stars.

\begin{figure*}
\epsfxsize=13cm
\leavevmode
\epsffile{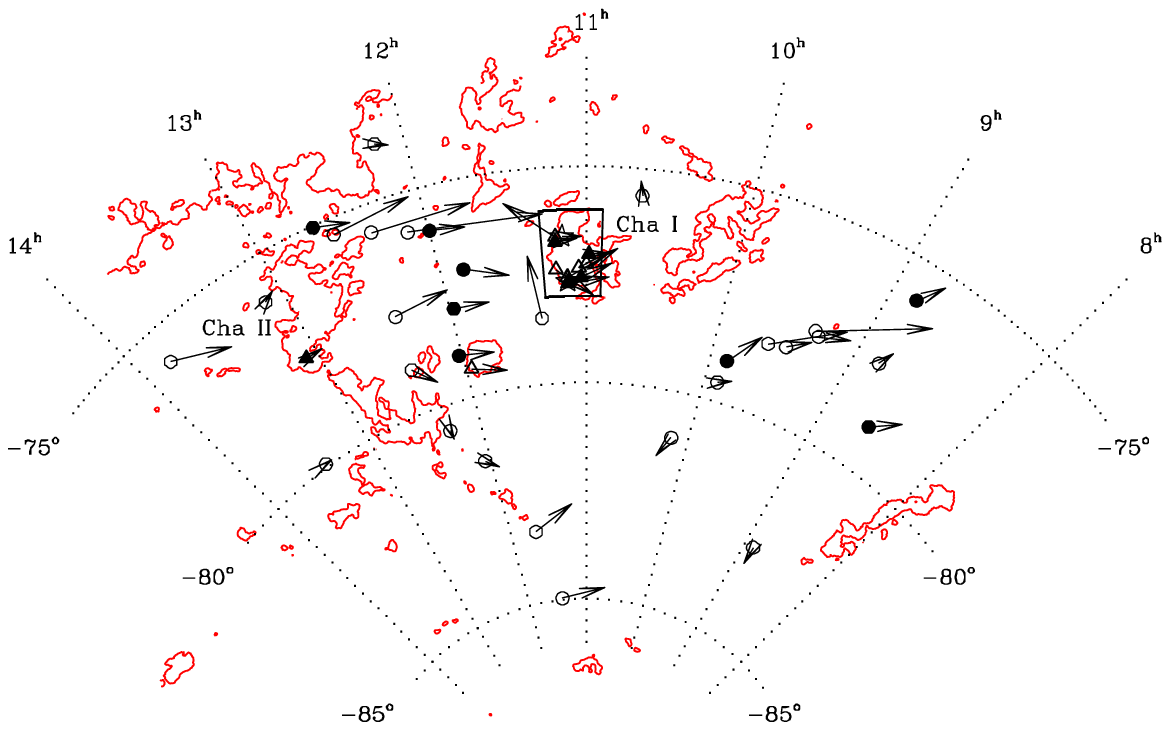}
%\hfill
%\parbox[b]{5.5cm}{
%\end{figure*}
%\begin{figure}[h]
\epsfxsize=5cm
\leavevmode
\epsffile{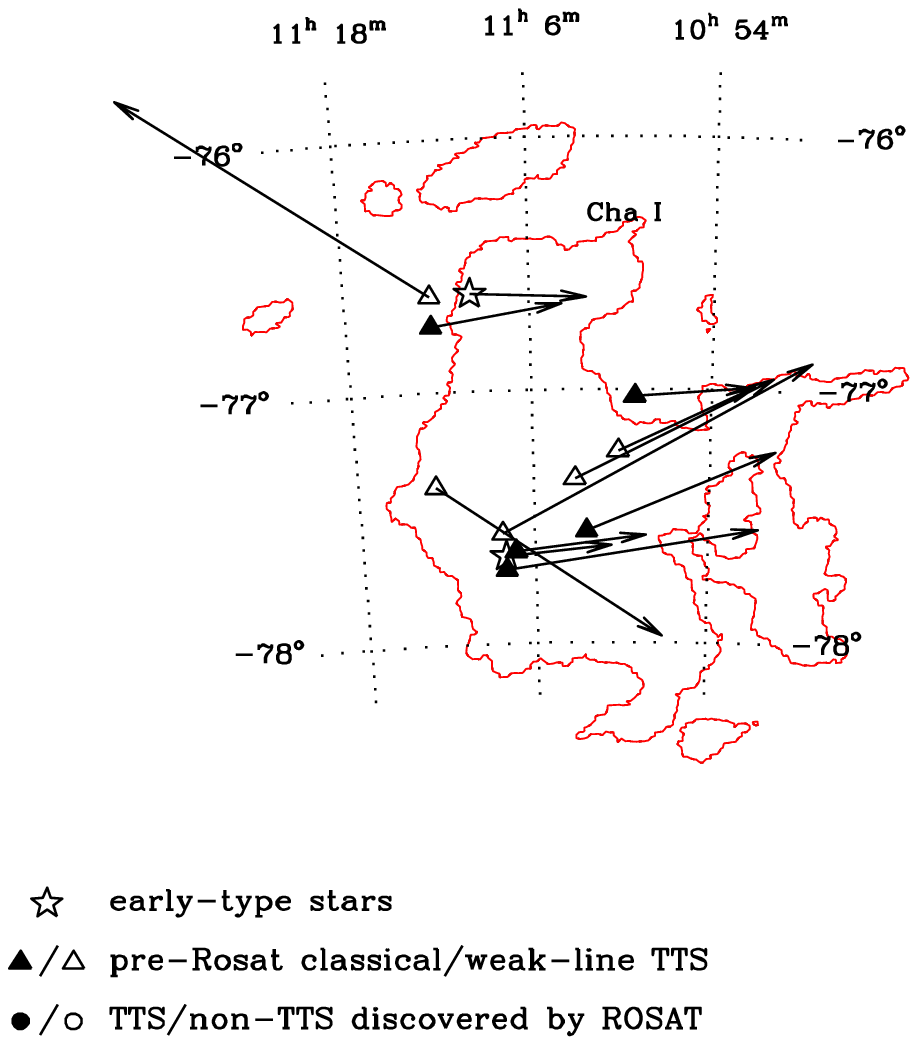}
\caption[]{\label{polar} Positions and proper motions of
the stars in Tables~\ref{liste1} and \ref{liste2}.
Contours are from the IRAS 100\,$\mu$m survey.
The region around Cha~{\sc i} is shown on an enlarged scale, too.
Most of the new ROSAT discovered stars are either located
between these two clouds or west of Cha~{\sc i}. 1\degr\ corresponds to 
50\,mas\,yr$^{-1}$; the largest arrow in the figure (RXJ~1207.9-7555
between Cha~{\sc i}
and Cha~{\sc ii}) corresponds to 156\,mas\,yr$^{-1}$.}
%}
\end{figure*}

%\begin{figure*}
%\resizebox{12cm}{!}{\includegraphics{polar.ps}}
%\hfill
%\parbox[b]{55mm}{\caption{   } label{polar}}
%\end{figure*}

A crucial point in this analysis are the individual distances of the
stars. Hipparcos parallaxes are available only for a very small fraction of 
our sample. The two bright late B-type stars HD\,97300 and
HD\,90480 known to be assoicated with the Cha\,{\sc i} cloud 
(Whittet et al.\ 1997) are located at distances of 188\,pc and 175\,pc, 
respectively. T~Cha seems to be located closer (66\,pc) than the other T~Tauri 
stars associated to Cha\,{\sc i} and Cha\,{\sc ii}, although the Hipparcos 
parallax has a very large error.
Sz\,6 is located at 143\,pc, and the Hipparcos results for Sz\,19 
and CV~Cha are uncertain.
For stars not measured by Hipparcos we adopt a mean value of 170\,pc unless
stated otherwise, taking the Hipparcos results (Wichmann et al.\ 1998) as well
as determinations based on various other methods (see Schwartz 1991 for a
review) into account. Note that this value is also in good agreement with
the recent distance estimate of 160$\pm$15\,pc to the Cha\,{\sc i} cloud by 
Whittet et al.\ (1997) derived on the basis of reddening distributions.
We make no distinction between the distance to the Cha\,{\sc i} and the
Cha\,{\sc ii} clouds, because indications for a larger distance to Cha\,{\sc ii}
are rather uncertain (Schwartz 1991, Brandner \& Zinnecker 1997, 
Whittet et al.\ 1997).

The paper is organized as follows: In Section~2 we present and discuss
our data, taken from several proper motion catalogues, and define
the samples. Section~3 is devoted to the kinematics of these stars;
proper motions and space velocities are investigated in detail.
Finally, we discuss the implications of these motions for several
star formation scenarios in Section~4 and present our conclusions in
Section~5.

\section{Data}

\subsection{Proper motion catalogues}
Astrometric data with the highest currently available accuracy 
is provided in the Hipparcos Catalogue (ESA 1997). 
It contains about 120\,000 stars and 
the typical error of the proper motions is about 1\,mas\,yr$^{-1}$.
The Hipparcos Catalogue was the major output
of the ESA Hipparcos space astrometry satellite mission, and proper motions
were determined by fitting all astrometric parameters (positions,
proper motions, parallax) simultaneously to the data points collected over
the about 3\,years time of operation for every individual star.

Proper motions in the ACT Reference Catalogue (Urban et al.\ 1997) and
STARNET (R\"oser 1996) however were determined by
comparing the positions of stars with an epoch difference of about 80\,years.
Both catalogues use the Astrographic Catalogue (AC) with a mean epoch of
1907 as the first position measurement. For the ACT the Tycho Catalogue
(ESA 1997) provides the second epoch, yielding proper motions for about
1 million stars with an accuracy of about 3\,mas\,yr$^{-1}$.
STARNET uses the Guide Star Catalogue (GSC 1.2) as second epoch and provides
proper motions for 4.3 million stars with an accuracy of about 
5\,mas\,yr$^{-1}$. Thus it is the most extensive proper motion catalogue 
available so far, containing stars with magnitudes up to 12\,mag and 
mean errors still acceptable for kinematic studies.

The proper motions discussed in the following sections are taken from these
three catalogues, which are all on the ICRS astrometric system defined by
Hipparcos. The PPM proper motions, with a similar accuracy as the ACT for
400\,000 stars, were used for comparison only, see Sect.~\ref{pmdiff}.

\subsection{The samples}
In Table~\ref{liste1} we list all
the stars which were known or suspected to be connected to the Chamaeleon
association before the ROSAT mission, along with their proper motions.
Altogether these are 14~stars: the 2 well-known late B type stars 
HD~97048 and HD~97300, with entries from Hipparcos,
and 6 classical and 6 weak-line T~Tauri stars. We have included T~Cha,
although it maybe located foreground to the Chamaeleon clouds as indicated by
its Hipparcos parallax (Wichmann et al.\ 1998). Only one of these stars 
(BF~Cha) is associated with the Cha~{\sc ii} cloud, whereas the other stars 
are located close to Cha~{\sc i}. There is a third cloud in the Chamaeleon 
region termed Cha~{\sc iii} which apparently also shows star formation activity 
(Pfau et al.\ 1996), but the sources are on average 2\,mag fainter than in 
the other two clouds and none could be identified in STARNET.

As in other nearby star forming regions, optical follow-up observations of 
X-ray sources discovered by ROSAT led to the identification of 77~probable 
new pre-main sequence stars in the Chamaeleon region (Alcal\'a et al.\ 1995,
1997). These new T~Tauri stars are not only located close to the clouds like
most of the T~Tauri stars known before, but also up to 10$^{o}$ away from any
known site of star formation.

Precise determinations of the lithium line (670.7\,nm) strength by means 
of high resolution spectroscopy and comparison with the typical lithium 
equivalent width of young main sequence stars (like the Pleiades) of the 
same spectral type confirmed the 
pre-main sequence nature for more than half of these stars (C97).
We could identify 31~stars of the total sample in the Hipparcos, ACT 
and STARNET catalogues (Table~\ref{liste2}).

Unfortunately, of the 31 stars newly discovered with ROSAT for which 
we can find proper motions only 8 are confirmed low-mass PMS stars, whereas
in the whole sample of C97 the fraction of bona-fide PMS to non-PMS stars 
is about twice (40 out of 81).
This is a consequence of the fact that most of the confirmed
low-mass PMS stars in the C97 sample having spectral types later than G5 are 
normally fainter than about $V\approx$\,11.5 mag and hence are not
included in the Hipparcos, ACT and/or STARNET catalogues, while the other
objects classified as ZAMS stars or with dubious PMS nature by C97 have
on average earlier spectral types and hence are sufficiently bright to
be present in the aforementioned catalogues.

\begin{figure}
\epsfxsize=8.8cm
\leavevmode
\epsffile{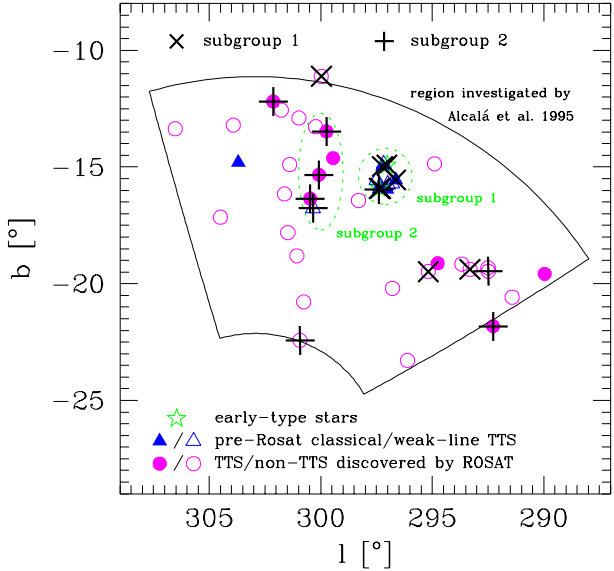}
\caption[]{\label{pos} Positions of the stars in Tables~\ref{liste1} and
\ref{liste2} in galactic coordinates. 
The subgroups~1 and 2 defined in Section~\protect\ref{subgroups} are
indicated by the overplotted symbols 'x' and '+'. Stars with dubious PMS nature
according to C97 and others which could be attributed to one of the
subgroups on the basis of their proper motions are also coded.
For illustration the approximate positions of subgroups~1 \& 2 on
the sky and the region investigated by Alcal\'a et al.\ (1995) are indicated,
too, whereas for subgroup~3 there is no pronounced clustering in the
position diagram.}
\end{figure}

Similarly, only sources detected with the ROSAT All-Sky Survey and
none of the sources detected only in ROSAT PSPC pointed observations
could be identified in any of the proper motion catalogues. This
means that in our sample there is no artificial spatial clustering due to 
possibly locally varying sensitivities present within the region indicated 
in Fig.~\ref{pos}.

\subsection{Discordant proper motions}
\label{pmdiff}
For the majority of the stars in Tables~\ref{liste1} and \ref{liste2} more
than one proper motion measurement is available, so that we are able to
compare its values in different catalogues. For most of our stars we find
no significant differences and we list the most accurate
determination.

\begin{table}
\caption[]{\label{liste3} Stars with discordant proper motions in one or
more catalogues. All proper motions have been transformed to the astrometric
reference system defined by Hipparcos, so no systematic differences should be
present.}
\begin{tabular}{lrrrr}
\noalign{\smallskip}
\hline\noalign{\smallskip}
[mas\,yr$^{-1}$] & 
%\multicolumn{1}{c}{\raisebox{1ex}[-1ex]{
$\mu_{\alpha}$ &
%}\raisebox{-1ex}[1ex]{\scriptsize \hspace*{-0.6cm}$\cos\delta$}} &
$\mu_{\delta}$ & 
\multicolumn{1}{c}{\raisebox{1ex}[-1ex]{$\sigma_{\mu_{\alpha}}$}
\raisebox{-1ex}[1ex]{\scriptsize \hspace*{-0.7cm}$\cos\delta$}} &
$\sigma_{\mu_{\delta}}$ \\
\noalign{\smallskip}
\hline\noalign{\smallskip}
\hline\noalign{\smallskip}
\multicolumn{5}{c}{Sz 19} \\
\hline\noalign{\smallskip}
HIP     & -113 &  3 & 3.1 & 2.9 \\
ACT     & -120 &  8 & 3.1 & 2.1 \\
STARNET & -106 & 45 & 4.4 & 4.5 \\
\hline\noalign{\smallskip}
\multicolumn{5}{c}{RXJ 0837.0-7856} \\
\hline\noalign{\smallskip}
ACT     & -147 & 26 & 1.4 & 1.8 \\
STARNET & -321 & 59 & 3.7 & 3.7 \\
\hline\noalign{\smallskip}
\multicolumn{5}{c}{RXJ 1125.8-8456} \\
\hline\noalign{\smallskip}
HIP     & -545 & 12 & 0.7 & 0.6 \\
PPM     & -482 & 12 & 2.1 & 2.4 \\
ACT     & -555 & 12 & 0.8 & 1.3 \\
STARNET & -388 &  9 & 3.2 & 3.2 \\
\hline\noalign{\smallskip}
\multicolumn{5}{c}{RXJ 1159.7-7601} \\
\hline\noalign{\smallskip}
HIP     & -165 & -5 & 1.7 & 1.5 \\
ACT     & -199 & -7 & 8.7 & 2.9 \\
STARNET & -224 &-20 & 3.8 & 3.8 \\
\noalign{\smallskip}\hline
\end{tabular}
\end{table}

Stars with discordant proper motions in two or more catalogues are listed
together with all available proper motion determinations in
Table~\ref{liste3}. The most probable reason for differences in the proper
motions are non-resolved binary or multiple systems: in general it is not clear
whether the photocentre or the brighter component was observed, and sometimes
this may be different for the positions of the first and second epoch,
especially for variable stars. This may lead to spurious proper motions in the
ACT and STARNET catalogues.

Orbital motion further complicates the determination of the mean proper 
motion for the whole system. The largest effect is expected for the Hipparcos
proper motions, since 3~years of data collection covers only a short fraction
of the orbits of long period binaries. Thus the instantaneous motion 
of the photocentre seen by Hipparcos does not reflect the mean motion of the
centre of mass for these kind of systems (Lindegren et al.\ 1997; Wielen 1997).

Two stars of Table~\ref{liste3} are present in the Double and Multiple Systems
Annex of the Hipparcos Catalogue, where the observational effects of duplicity
have been taken into account. Sz~19 is perhaps an
astrometric binary with a short period which could not be resolved by
Hipparcos. Indeed Schwartz (1977) notes a close companion to Sz~19 in the 
south, confirmed by Reipurth \& Zinnecker (1993) and Ghez et al.\ (1997), and 
the secondary is variable with magnitude differences of at least 2.5\,mag 
(Brandner 1992).
For RXJ~1125.8-8456 a non-linear model of the motion including acceleration
terms was fitted to the Hipparcos observations, which has no meaning outside the mission
interval. Although this is formally a single-star solution, we may deal with an
unresolved system with a period in the range 10-100\,years (Lindegren et 
al.\ 1997).

\begin{figure}
\epsfxsize=8.8cm
\leavevmode
\epsffile{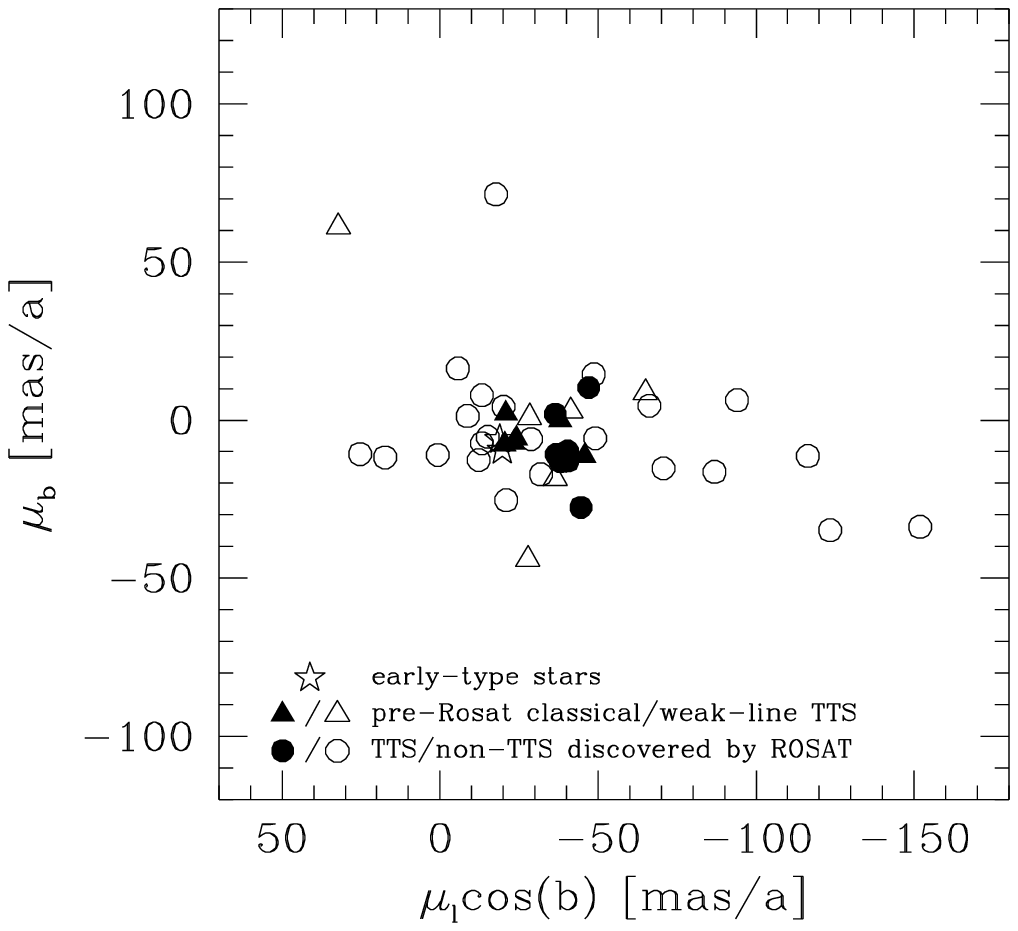} \\
\epsfxsize=8.8cm
\leavevmode
\epsffile{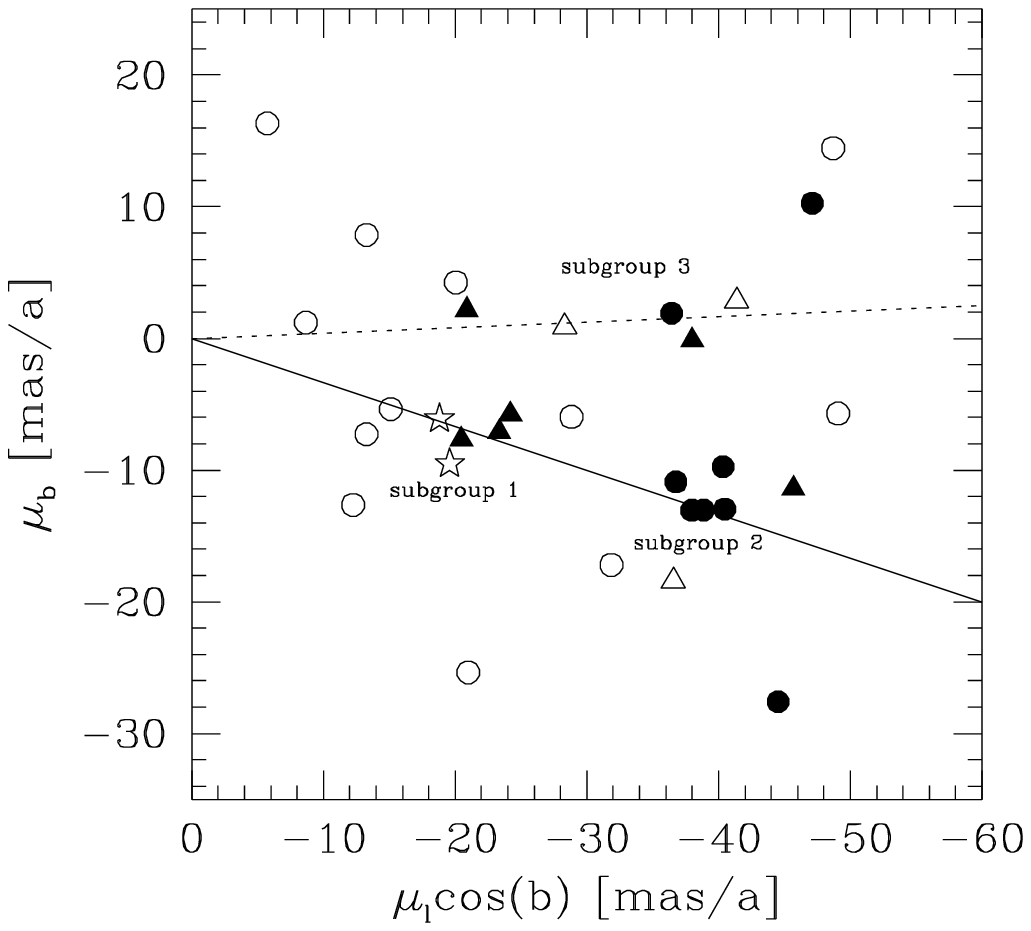}
\caption[]{\label{pm} Proper motion diagram in galactic coordinates for all the
stars in Tables~\ref{liste1} and \ref{liste2}. The lower panel is an
enlarged reproduction of the center of the upper panel. 
The different subgroups are introduced in Section~\protect\ref{subgroups}, 
and the lines correspond to motions for varying distances.}
\end{figure}

Similar effects may be responsible for the inconsistencies in the proper
motions of the other two stars from Table~\ref{liste3}:
RXJ~0837.0-7856 was not observed by Hipparcos, but is flagged as dubious
astrometric reference star in the Tycho catalogue, as is RXJ~1159.7-7601. The
latter star is additionally flagged as 'perhaps non-single' in the ACT
catalogue. 

Another source of errors in the ACT and STARNET catalogues may be wrong
identifications of stars, favoured by large epoch differences and large 
proper motions.
All these effects may explain the large differences between proper motions in
different catalogues, although in most of the cases the actual error source is 
difficult to find out.

\section{Kinematics}

\subsection{Proper motions}
Positions and proper motions of the stars in Tables~\ref{liste1} and
\ref{liste2} are plotted in Fig.~\ref{polar}.
The two open star symbols denote the early type stars HD~97048 and
HD~97300, the filled triangles classical TTS and the
open triangles weak-line TTS known before the ROSAT mission.
The filled circles represent confirmed low-mass PMS stars, while the open
ones are objects classified as ZAMS stars or with dubious PMS nature by C97.
One immediately notes that
there is a trend for proper motion vectors pointing to the west, with some
scatter especially for the ZAMS stars, as expected from their probably
higher velocity dispersion. For a more detailed
analysis, we also plot the data in galactic coordinates (Figs.~\ref{pos} and
\ref{pm}), which are better suited for a rectangular illustration due to the
position of the Chamaeleon association near to the southern equatorial pole.
The motion of the stars is partly due to the reflex motion of the Sun,
which is about ($\mu_{l}\cos b,\mu_{b}$)$\approx$ ($-17$,$-6$)\,mas\,yr$^{-1}$
at the position of the Chamaeleon association and a distance of 170\,pc.
Note however that this value strongly depends on the adopted distance
(for half the distance the value would be twice as high), whereas
the variations caused by different positions on the sky are rather low 
within our study area.

\begin{table}
\caption[]{\label{sub} Subgroups derived from the proper motion diagram
(Fig.~\protect\ref{pm}) with their mean proper motions and dispersions. 
The number of stars in each subgroup is given in the last column.}
\begin{tabular}{cr@{$\;\pm\;$}rr@{$\;\pm\;$}rc}
\hline\noalign{\smallskip}
sub- &
\multicolumn{2}{c}{$\mu_{l} \cos b$} &
\multicolumn{2}{c}{$\mu_{b}$} & 
\# of \\
group & \multicolumn{2}{c}{[mas yr$^{-1}$]} &
\multicolumn{2}{c}{[mas yr$^{-1}$]} & stars \\
\noalign{\smallskip}\hline\noalign{\smallskip}
1 & -21.3 & 2.4 &  -7.2 & 1.5 & 5 \\
2 & -39.5 & 3.1 & -12.8 & 2.8 & 7 \\
3 & -38.6 & 2.5 &   1.6 & 1.5 & 3 \\
\noalign{\smallskip}\hline
\end{tabular}
\end{table}

\subsubsection{Bona-fide PMS stars}
\label{subgroups}
From Fig.~\ref{pm} we infer at least 2 or 3 different areas 
in the proper motion diagram where confirmed PMS stars tend to 
cluster (Table~\ref{sub}).
It turns out that these subgroups are not only apparent in the proper motion 
diagram, but likewise they correspond to different regions in the
position diagram, which independently confirms our subdivisions.
The first subgroup consists of the 2~early type stars and 3~CTTS (Sz~6,
Sz~19 and CV~Cha), all located in the cloud core of Cha~{\sc i}.
In the second subgroup there are 5~new PMS stars, which all have very similar
proper motions and, with the exception of RXJ~0837.0-7856\footnote{Note 
from Table~\protect\ref{liste3} that RXJ~0837.0-7856 has a
different proper motion in STARNET which would make it a good candidate for a
run-away TTS (RATTS),
possibly ejected some $10^{6}$\,years ago.}, 
are all located between the Cha~{\sc i} and Cha~{\sc ii} clouds.
Besides these ROSAT detected PMS stars one CTTS (VW~Cha) and one 
WTTS (T~Cha) also
match the requirements of subgroup~2. Note that T~Cha is also located between
Cha~{\sc i} and Cha~{\sc ii}, whereas the position of VW~Cha is close to
the core of the Cha~{\sc i} cloud.

\begin{table*}
%\begin{center}
\caption[]{\label{disp} Mean space velocities and dispersions
separately for stars of subgroups~1 \& 2 and Hipparcos PMS stars as well
as for the total samples, as shown in Fig.~\ref{uvw}.
$U$-velocities are positive in the direction of the galactic centre.
The velocities have been corrected for the effects of differential 
galactic rotation and the reflex motion of the Sun.
For the calculation of the space velocities we used either the Hipparcos
parallaxes or, where not available, a distance of 170\,pc for stars in
subgroup~1 and 90\,pc for stars in subgroup~2.}
\begin{tabular}{l@{$\qquad$}r@{$\qquad$}cr@{$\qquad$}cr@{$\qquad$}cr}
\hline\noalign{\smallskip}
& & \multicolumn{1}{c}{$<U>$} & \multicolumn{1}{c}{$\sigma_{U}$} & 
\multicolumn{1}{c}{$<V>$} & \multicolumn{1}{c}{$\sigma_{V}$} &
\multicolumn{1}{c}{$<W>$} & \multicolumn{1}{c}{$\sigma_{W}$} \\
 & \raisebox{1.5ex}[-1.5ex]{\#}& 
\multicolumn{2}{c}{[km\,s$^{-1}$]} & \multicolumn{2}{c}{[km\,s$^{-1}$]} &
\multicolumn{2}{c}{[km\,s$^{-1}$]} \\
\noalign{\smallskip}\hline\noalign{\smallskip}
subgroup 1            &  4 & \hphantom{-1}4.4 & 6.7 & -11.0 & 8.1 & -3.4 & 2.7 \\
subgroup 2            &  7 & \hphantom{-1}3.1 & 2.8 & \hphantom{1}-7.6 & 3.3 & -2.3 & 1.6 \\
subgroups 1 \& 2      & 11 & \hphantom{-1}3.6 & 4.3 & \hphantom{1}-8.8 & 5.4 & -2.7 & 2.0 \\
Hipparcos PMS         &  6 & \hphantom{-1}5.2 & 5.5 & \hphantom{1}-9.6 & 6.7 & -3.1 & 2.5 \\
all Hipparcos stars   & 11 & -14.9 & 30.6 & -10.7 & 18.5 & -1.1 & 11.9 \\
\noalign{\smallskip}\hline\\[0.8ex]
\end{tabular}
%\end{center}
\end{table*}

\begin{figure*}
\epsfxsize=12cm
\leavevmode
\epsffile{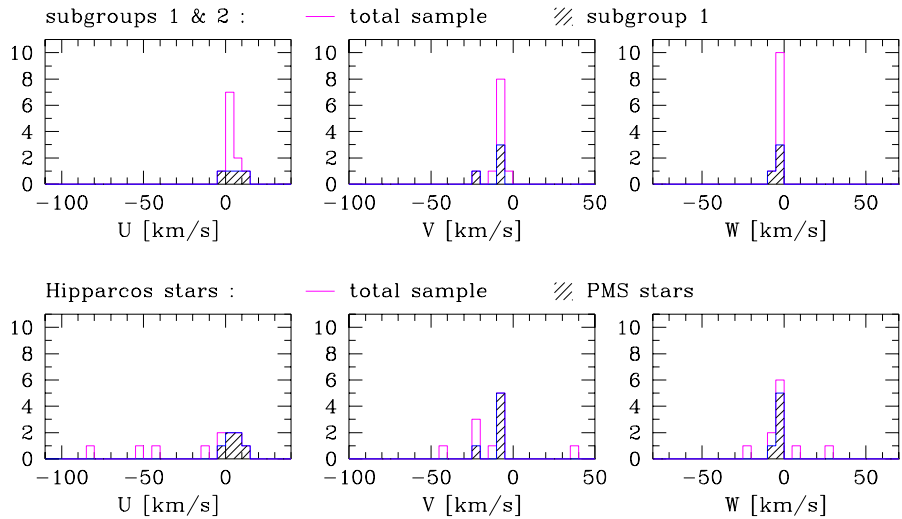}
\hfill
\parbox[b]{5.5cm}{\caption[]{\label{uvw} Space velocity histograms shown
separately for stars in subgroups~1 and 2 (upper panels) and for
Hipparcos stars (lower panels). In the upper diagrams the contribution
of subgroup~1 is hatched, and no significant differences between the
two subgroups are visible. In the lower diagrams the contribution of the
PMS stars is hatched. 
We assumed the same distances as in Table~\ref{disp} and corrected the
velocities for the effects of differential galactic rotation and the motion
of the Sun.}}
\end{figure*}

The proper motions of subgroups~1 and 2 point into the same direction,
but the absolute values are about twice as high for the second group.
This finding is consistent with a scenario where the mean distance 
for the second subgroup is about half of
the mean distance for the first subgroup. Then, both groups would have 
consistent space velocities.
Indeed this picture is confirmed by the Hipparcos parallaxes: with our
assumption for the mean distance of Cha~{\sc i} of 170\,pc (the 
distances for
4~stars in subgroup~1 are 175\,pc (HD~97048), 188\,pc (HD~97300), 143\,pc
(Sz~6) and 210\,pc (Sz~19)\,) we would expect a mean distance of about
90\,pc for stars in our subgroup~2. The parallaxes as observed by Hipparcos for
3~stars in subgroup~2 correspond to distances of 
86\,pc (RXJ~1158.5-7754a), 92\,pc (RXJ~1159.7-7601) and 66\,pc (T~Cha), 
which gives very strong support to our interpretation.

The existence of subgroup~3 is not so obvious as for the other
2~subgroups. The 3 stars which we grouped together are CS~Cha, CHXR~11
and RXJ~0850.1-7554. Nevertheless, if we assume its existence we could 
attribute 4~more stars with higher (BF~Cha and CHXR~8) or lower distances 
(RXJ~0951.9-7901 and CHXR~32) to it.

Only 2~WTTS and 1~new PMS (Sz~41, CHXR~56 and RXJ~1150.4-7704) are left
from the sample of the bona-fide PMS stars (Table~\ref{liste1} and upper part
of Table~\ref{liste2}) which
do not fit in any of the above subgroups because of quite different proper 
motions. Sz~41 is at least a double system: besides a faint companion another 
star nearly as bright as the primary is located 11.4$\arcsec$ away from Sz~41 
(Brandner 1992; Reipurth \& Zinnecker 1993). RXJ~1150.4-7704 is flagged
as a possible spectroscopic binary by C97. 
Thus it is possible that the proper motions of these stars are not
representing their space motions.

The velocity dispersions in our subgroups are of the same order of
magnitude as the errors of the proper motions, and so the intrinsic
velocity dispersions must be much smaller
(at a distance of 170\,pc 1\,mas\,yr$^{-1}$ corresponds to 0.8\,km\,s$^{-1}$).
To some extent this was expected, because we only grouped stars with similar
proper motions together. On the other hand such low values for the intrinsic
velocity dispersion agree with other determinations. Jones \& Herbig (1979)
derived a value of 1-2\,km\,s$^{-1}$ in one coordinate for the
intrinsic velocity dispersions of subgroups in Taurus-Auriga and 
considered this as typical for associations. Dubath et al.\ (1996)
calculated a value of 0.9$\pm$0.3\,km\,s$^{-1}$ based on the radial 
velocities of 10 stars in Cha~{\sc i}.

\subsubsection{ZAMS stars and others}

The stars of Table~\ref{liste2} classified as stars with dubious PMS nature
or as ZAMS stars by C97 clearly show a very large range in proper motions, 
which independently confirms the conclusions from applying the lithium
criterion by C97. This criterion is very conservative, as it rejects
stars with lithium abundances similar to the Pleiades as weak-line TTS,
although it may very well be the case that some truly pre-main sequence stars
exhibit such low lithium strength.
Note that all the stars with Hipparcos parallaxes in Table~\ref{liste2} 
fall well above the main sequence when comparing their positions in the
HR diagram with various PMS evolutionary tracks (Neuh\"auser \&
Brandner 1998).

There are a few other stars in Table~\ref{liste2} which - rated from their
proper motions - probably fall into this category of unrecognized weak-line
T~Tauri stars. Judging from the proper motions alone one could
assign RXJ~0928.5-7815, RXJ~0952.7-7933 and RXJ~1209.8-7344 to
subgroup~1, RXJ~0917.2-7744 and RXJ~1125.8-8456 (its Hipparcos parallax
corresponding to
83\,pc also fits this interpretation) to subgroup~2, and
RXJ 0849.2-7735 and RXJ~1223.5-7740 to subgroup~3.
One must however bear in mind that the reflex motion of the sun is very similar
to the typical proper motion of Cha\-mae\-leon member stars, making it difficult
to distinguish between members and field stars on the basis of the proper
motions alone.

\subsection{Space velocities}
\label{space}
We have calculated space velocities for all stars in subgroups~1 and 2 and
the Hipparcos stars with radial velocities available in the literature
(HD~97048 from Finkenzeller \& Jankovics (1984), 4~CTTS from
Dubath et al.\ (1996), and T~Cha and stars in Table~\ref{liste2} 
from C97).
For stars not observed by Hipparcos a distance of 170\,pc 
for subgroup~1 and 90\,pc for subgroup~2 was adopted.

We corrected the space velocities for the effect of differential galactic
rotation, assuming the IAU standard values of 8.5\,kpc for the distance
to the galactic centre and 220\,km\,s$^{-1}$ for the velocity of the
Local Standard of Rest.
The value of this correction  depends on the galactic azimuthal angle and
therefore in general also on the distances of the stars.
The mean corrections in the $U$-velocities for stars in subgroup~1 and 2 
are 3.8\,km\,s$^{-1}$ and 2.0\,km\,s$^{-1}$, respectively (the corrections
in the $V$-velocities are practically zero).
Additionally, the motion of the Sun (\,(U,V,W) = (9,12,7)\,km\,s$^{-1}$,
Delhaye 1965) has been added to the space velocites, although it does not
change the relative velocities between the groups which are of interest 
here.

The coincidence of the mean values for the three space velocity 
components of Hipparcos PMS stars and the combined sample of 
subgroups~1 \& 2 is artificial to some extent (see Table~\ref{disp})
as the 6 Hipparcos stars form a subset of subgroups~1 \& 2.
The mean values for subgroup~1 and 2 are also in quite good agreement.

Our interpretation of different proper motions in terms of different distances
is further confirmed
when taking the additional information on the radial velocities and projection
effects due to different positions in space into account.

Comparing the Hipparcos PMS stars with ZAMS stars and stars of dubious PMS
nature in Fig.~\ref{uvw}, one again notes the clear peak in the distribution
of the PMS stars and the large scatter of the presumably older stars.

\section{Discussion}

\subsection{Structure of the Chamaeleon clouds}
The IRAS 100$\mu$m map of the Chamaeleon region shows several
filamentary clouds which extend over an area of more than 100 square 
degrees. It is an open question whether the individual structures
termed Cha~{\sc i}, Cha~{\sc ii} and Cha~{\sc iii} are really
physically related to each other. There is another cloud, DC~300.2-16.9
(Hartley et al.\ 1986),
located between Cha~{\sc i} and Cha~{\sc ii} roughly at the position
of T~Cha. 

The Hipparcos parallax of T~Cha implies a relatively small distance of
66\,pc. Note however that the large parallax error for this star puts an
upper limit of 85\,pc while the lower limit is 54\,pc. The latter would
place T~Cha practically on the main sequence, which is absolutely
inconsistent with the pronounced PMS characteristics of this star
(Alcal\'a et al.\ 1993). Even the mean distance of 66\,pc would give an
extremely old age of some 40\,Myr for this star. Since T~Cha is definitely
a T~Tauri star, we think that the upper limit of 85\,pc should be closer to
the true distance of T~Cha.

On the other hand, an upper limit of 180\,pc has been established for
the distance of the cloud DC~300.2-16.9 (Boulanger et al.\ 1998),
to which T~Cha seems to be associated.
Thus, it may well be that this cloud is also located closer than the
Cha~{\sc i} cloud, maybe also at about 90\,pc from the Sun.
The Hipparcos parallaxes of the other stars in subgroup~2 as well as
our analysis of the proper motions in Section~\ref{subgroups} support
this scenario of stars and even some cloud material at distances of
about 90\,pc.

The question now is how can a SFR be such large in volume?
We discuss the models for the formation scenario of the Chamaeleon
cloud complex which possibly could explain the existence of PMS stars
far off the observed molecular clouds in the next section.

Alternatively, we may note that it is also possible that the observed
cloud material belongs to distinct structures as considered by 
Whittet et al.\ (1997).
In this case the stars in subgroups~1 \& 2 would have the same space
velocities although they are not associated with the same cloud material.
However, it is not unusual that young stars exhibit rather low velocities
relative to the field stars in the same region (cf.\ the Taurus SFR or
the Scorpius-Centaurus OB association).

Moreover, placing the stars of subgroup~2 at a mean distance of 90\,pc rises
their mean ages by about a factor of 6 to 18~Myr as compared to a mean
age of 3~Myr for a mean distance of 170\,pc. This could easily be explained
if they belong to another structure than the stars of subgroup~1.

There are too few stars in the Chamaeleon region with distance information
available to decide whether a population of PMS stars with distances
intermediate between the two subgroups at around 130\,pc exists.
In principle, the majority of dispersed T~Tauri stars detected with
the flux-limited RASS are expected to be located between 90\,pc and
150\,pc, and the optical, IR and deep X-ray pointed observations have been
sufficiently sensitive to detect PMS stars at more than 150\,pc.
However, most of these stars were too faint to be included in the
Hipparcos Input Catalogue.

\subsection{Implications for the formation scenario of the Chamaeleon 
cloud complex}
The discovery of large populations of WTTS distributed over regions of 
10-20~degrees or even more in extent centered around active cloud cores has 
raised questions about the scenario of their formation.
Several scenarios have been suggested to account for the existence of very 
young stars far away from the known sites of star formation.

Sterzik \& Durisen (1995) proposed that the WTTS halo observed around
star formation regions might be due to
high velocity (\,$\ga$\,3\,km\,s$^{-1}$\,) escapers (run-away TTS or
RATTS) produced by dynamical interactions in small stellar systems.
This would imply that
the velocity vectors of the stars point away from the dense molecular 
cloud cores from where they were ejected.

From our proper motion study there is no indication for such an overall
correlation between positions and proper motions. Only for some 2 or
3~stars in our sample the ejection scenario may be invoked, namely
Sz~41, CHXR~56, and perhaps RXJ~0837.0-7856, if their proper motions are
not spurious due to binarity.
On the contrary, if we assume that subgroup~2 is at the same distance
as subgroup~1 (ignoring the Hipparcos parallaxes for the moment), the
stars of subgroup~2 would move with higher space velocities in the direction of
lower right ascension than the stars of subgroup~1
while being located at higher right ascension (cf.\ Fig.~\ref{polar}).
This means that they would approach the Cha\,{\sc i} molecular cloud.
Given the direction of motion, we cannot exclude that some stars may have
been ejected from the Cha\,{\sc ii} cloud. However, the fact that these stars
seem to form a co-moving group is inconsistent with the prediction of any
ejection model, in which the motion would be completely random, so that we
exclude the ejection scenario as the dominant process for producing the
dispersed population of WTTS in Chamaeleon.

L\'epine \& Duvert (1994) tried to explain the displacement with
respect to the galactic plane of several near-by star forming regions 
including Chamaeleon by infall of high velocity clouds (HVC) on the
galactic plane. There is no detailed prediction for the kinematics of
the stars in the HVC impact scenario, except for the fact that,
subsequent to the impact, clouds and stars will oscillate around the
galactic plane and tend to separate from each other (combing-out).
Given the fact that at least the stars in subgroups~1~\&~2 display
practically no
net motion perpendicular to the galactic plane after correcting their
proper motions for the solar reflex motion, one might speculate that they are
just reversing their direction of motion. The large distance of
the Chamaeleon association from the galactic plane may
support this point of view. Nevertheless, these indications are far
from being conclusive and depend strongly on the adopted distances.

Feigelson (1996) proposed that low mass stars may form in dispersed 
cloudlets in a turbulent environment. Also, it has long been suspected
that Bok globules are the sites of isolated star formation.
Recent studies (Launhardt \& Henning 1997, Yun et al.\ 1997) 
have shown that such globules can be
associated with embedded IR and IRAS point sources in which very young
low mass stars are found. It is however not clear if these globules are
related to Feigelson's cloudlets.

In order to explain the observed distribution of WTTS Feigelson (1996)
considers models with a velocity dispersion of the order of 1\,km\,s$^{-1}$
(due to internal thermal motions in the gas of the parent cloud), 
with thermal velocity dispersal in combination with dynamical ejection, and 
with star formation in small cloudlets distributed over a larger region.
Comparing the predictions of his models with the properties of the observed 
WTTS population, he found that the first two dispersal models encounter
serious problems. In particular,
the thermal dispersal model can explain the number of WTTS found far
from the active clouds, but not their low ages. 
In order to overcome this
difficulty within the framework of the current model one would have to
assume an unplausibly high velocity dispersion even at the time of their
formation.
An improved dispersal model, where a certain fraction of the
dispersing stars is made up of high velocity escapers, cannot account
for the observations either, unless the ejection rate significantly
increased recently. The model can indeed explain the existence of
some very young stars far from their sites of origin, but
simultaneously it produces a population of older ejected stars, which
would lead to an unplausibly high star formation efficiency.

As already pointed out above, the proper motion data of the stars
discussed in the previous sections are inconsistent with any dispersal
model either, as the velocity vectors are not oriented away from any
single point.

In the most promising model investigated by Feigelson star formation
takes place in long-lived active cloud cores as well as in a number of
small short-lived cloudlets distributed over a rather large
region. These cloudlets are believed to possess high velocities
relative to their parental giant molecular clouds because of its turbulent
structure. After producing some stars with very low internal velocity
dispersion the cloudlets disappear, leaving behind streams of T~Tauri
stars with high relative velocities between each other.

As proper motions are available only for a very small fraction of
all the young stars in the Chamaeleon region (probably for less than
10\% according to the estimate by Feigelson of several hundred stars
yet to be discovered), it is difficult to verify the predictions of the
model quantitavely. If we ignore the stars in subgroups~1~\&~2 for the
moment, which we assume to have formed in the dense cloud cores,
we are left with not more than 33~stars which possibly 
originated in small cloudlets. Feigelson estimates the number of
cloudlets to be of the order of 50, so that we do not expect to have
more than one or two stars of the same cloudlet in our sample.
Although we cannot confirm the model decisively, from the proper motion
diagram (Fig.~\ref{pm}) one could select good candidate stars 
which possibly were formed in cloudlets.

Another limiting factor in our kinematical study is the lack of 
precise distances for the majority of our stars. For a more detailed
comparision with the model of Feigelson one needs to correct the proper
motions of the wider distributed TTS and not only of the stars in the
two subgroups for galactic rotation and the reflex of the solar motion,
which requires knowledge of the individual distances.
Similarly, comparisions with the ejection scenario are also hampered by
the lack of precise distances, as the relative velocities can change
sign when putting the stars at higher or lower distances.

\section{Summary and conclusions}
We have analysed proper motions from the Hipparcos, ACT and STARNET
catalogues for altogether 45~stars, 22 of which are bona-fide pre-main
sequence stars and 23 are of dubious PMS nature or ZAMS stars.
On the basis of the distribution of the proper motions the presence of 
several subgroups in our data is suggested, which roughly coincide with 
similar groups on the sky. 
Given the kinematic distances which are independently confirmed by
the Hipparcos parallaxes, the two subgroups might belong to distinct
structures of the Chamaeleon clouds.

There is no indication in these data for a slow dispersal of stars out of
the active cloud cores, and so the model proposed by Sterzik \& Durisen 
(1995) cannot account for the large distribution of WTTS observed 
far from any known cloud material.
However, the observed motions are more or less consistent with the
high velocity cloud impact model of L\'epine \& Duvert (1994) if the
stars are currently at their turning point. Similarly, the data could be
interpreted in terms of the cloudlet model (Feigelson 1996) where star
formation takes place far off the active cloud cores and which
produces small groups of TTS with low internal velocity dispersions,
but high relative velocities between groups.

Larger proper motion catalogues produced hopefully by future space missions
like DIVA or GAIA will help to settle the question of the formation
scenario of this large population of weak-line T~Tauri stars in the 
Chamaeleon region.

\begin{acknowledgements}
We would like to thank the referee, Anthony Brown, for helpful suggestions
to improve the paper.\\
SF acknowledges grants from the Deutsche Forschungsgemeinschaft 
(DFG Schwerpunktprogramm `Physics of star formation').
WB acknowledges support under NASA/HST grant GO-07412.01-94A.\\
This research has made use of the SIMBAD database, operated at CDS,
Strasbourg, France.
\end{acknowledgements}

\end{document}